# Wavelength-scale noise-resistant on-chip spectrometer

Jianbo Yu[1,†], Hsuan Lo[2,†], Wenduo Chen[1], Changyan Zhu[2], Yujin Wu[1], Fakun Wang[1], Chongwu Wang[1], Congliao Yan[1], Cuong Dang[1], Bihan Wen[1], Hui Cao[3], Yidong Chong[2,4*], and Qi Jie Wang[1,2*]

[1]School of Electrical and Electronic Engineering, Nanyang Technological University, 639798, Singapore

[2]School of Physical and Mathematical Sciences, Nanyang Technological University, 637371, Singapore

[3]Department of Applied Physics, Yale University, New Haven, Connecticut 06520, USA

[4]Centre for Disruptive Photonic Technologies, Nanyang Technological University, 637371, Singapore

[†]The authors contributed equally to this work

* Corresponding authors: qjwang@ntu.edu.sg; yidong@ntu.edu.sg

**Abstract**

Performant on-chip spectrometers are important for advancing sensing technologies, from environmental monitoring to biomedical diagnostics. As device footprints approach the scale of the operating wavelength, previously strategies, including those relying on multiple scattering in diffusive media, face fundamental accuracy constraints tied to limited optical path lengths. Here, we demonstrate a wavelength-scale, CMOS-compatible on-chip spectrometer that overcomes this challenge by exploiting inverse-designed quasinormal modes in a complex photonic resonator. These modes extend the effective optical path length beyond the physical device dimensions, producing highly de-correlated spectral responses. We show that this strategy is theoretically optimal for minimizing spectral reconstruction error in the presence of measurement noise. The fabricated spectrometer occupies a lateral footprint of only 3.5 times the free-space operating wavelength, with a spectral resolution of 10 nm across the 3.59-3.76 μm mid-infrared band, which is suitable for molecular sensing. The design of this miniaturized noise-resistant spectrometer is readily extensible to other portions of the electromagnetic spectrum, paving the way for lab-on-a-chip devices, chemical sensors, and other applications.

**Introduction**

Optical spectrometry is essential to all scientific fields that touch upon the interactions of light and matter[1], and in recent years growing demand for compact integrated devices has driven the development of highly miniaturized spectrometers[2-12]. Spectrometers based on filtering[13,14], dispersive components[15-17], and Fourier transform interferometry[18,19] have been integrated into on-chip platforms, reducing the footprint required for spectrometry to several hundred operating wavelengths. A recent breakthrough has come from so-called reconstructive spectrometers[20-22], in which input spectral information is encoded in the intensities of space- or time-division output channels via a spectral transmission matrix, and retrieved with the help of reconstruction algorithms. Reconstructive spectrometers are able to resolve a broad set of spectra with only a few sampling channels[23,24], offering significant advantages in terms of compactness, application specificity, and cost-effectiveness.

A reconstructive spectrometer should ideally capture as much information as possible in the input signal, while being resistant to noise (including measurement noise in the photodetectors, environmental fluctuations, etc.). A common design strategy is to decrease the coherence of the spectral transmission matrix by raising the in-device optical path length, such as by leveraging multiple scattering and interference in disordered media[20]. In the diffusive scattering regime, the mean optical path length scales as $L^2/\ell_t$, where $L$ is the spectrometer size and $\ell_t$ is the transport mean free path. With strong scattering (which shortens $\ell_t$), it has been possible to achieve well-functioning disordered spectrometers with $L$ as small as tens of operating wavelengths, and mean optical path lengths substantially exceeding $L$.[21,25,26] This approach begins to fail, however, as the device size decreases toward the $L \sim \ell_t$ regime, at which point there is no path-length enhancement. A possible alternative is to make use of selected high-Q quasinormal modes, whose effective optical path lengths can be higher than the mean values predicted by diffusive scattering theory[27]. However, there are many open questions about this approach, including how best to engineer such quasinormal modes in a compact device, and how to optimize their properties for the purposes of spectral reconstruction.

In this work, we demonstrate a reconstructive spectrometer that operates robustly at a wavelength-scale footprint. The device's design is guided by a rigorous derivation of the optimal spectral transmission matrix that minimizes the reconstruction mean squared error (MSE) in the presence of noise, accompanied by the use of inverse design to engineer a photonic microcavity with quasinormal modes yielding a spectral transmission matrix of the desired form. Unlike approaches relying on optical diffusion, which is subject to the aforementioned scaling limits, we aim to couple specific individual modes to only one or a few output ports. These out-coupled modes have high effective optical path lengths, allowing for robust frequency discrimination even in a cavity on the order of a few wavelengths. As a proof of concept, we realize an on-chip spectrometer with a lateral footprint of only 3.5 times the free-space operating wavelength, an order of magnitude smaller than previous reported on-chip spectrometers. Our device achieves a spectral resolution of 10 nm over the 3.59-3.76 μm mid-infrared range, enabling the detection of molecules with mid-infrared absorption peaks such as formaldehyde and methane[28]. These results are obtained at a relatively high noise level of 7.9% in the output intensity, which is promising for device integration with low-power on-chip light sources. This work establishes a design paradigm for ultracompact, CMOS-compatible spectrometers, which can serve as the foundation for next-generation integrated photonic sensing platforms.

**Results**

*Design of a noise-resistant transmission matrix*

Our reconstructive spectrometer design is based on formulating a link between the error rate of spectral reconstruction and the scattering properties of a photonic structure. Consider a reconstructive spectrometer that encodes an incident spectrum, discretized across $n$ wavelengths as $\mathbf{s_0} = (s_1, s_2, \cdots, s_n)^T$, into the intensities of $m$ output channels $\mathbf{i_0} = (i_1, i_2, \cdots, i_m)^T$. With the inclusion of measurement noise, the detected intensities $\mathbf{i}$ are

$$\mathbf{i} = \mathbf{i_0} + \boldsymbol{\epsilon} = \mathbf{T}\mathbf{s_0} + \boldsymbol{\epsilon}, \qquad (1)$$

where $\mathbf{T}$ is the $m \times n$ spectral transmission matrix, and the vector $\boldsymbol{\epsilon}$ represents the random noise in each output channel. To minimize device footprint, one typically desires $m \ll n$, rendering Equation (1) an ill-posed problem. However, an approximate solution can be reconstructed using numerical techniques, such as a minimum L2-norm solution via the pseudoinverse[29]: $\mathbf{s} = \mathbf{T}^+\mathbf{i}$. Assuming that the measurement noise in each channel is independent and has the same standard deviation $\sigma$, the MSE has the following lower bound, whose derivation is given in the Supplementary Information:

$$E\{\|\mathbf{s} - \mathbf{s_0}\|_2^2\} \geq \sigma^2 \text{Tr}\{(\mathbf{T}\mathbf{T}^T)^{-1}\} \geq \frac{\sigma^2 m^2}{n}. \tag{2}$$

Reaching this lower bound requires $\mathbf{T}\mathbf{T}^T = \frac{n}{m}\mathbb{I}$, where $\mathbb{I}$ is the identity matrix. This condition can be decomposed into two requirements: (i) zero cross-correlation between transmission functions of each channel, and (ii) identical and maximized total transmission across all channels. This implies that the MSE can be minimized by having quasinormal modes that are each strongly coupled to one (or a few) output channels, as well as the spectral response in each channel having equal transmission and minimal spectral overlap. Based on these conditions, we define a figure of merit (FOM) for the reconstructive spectrometer consisting of the reciprocal of the trace of the middle bound in Equation (2):

$$\text{FOM} \equiv \text{Tr}\{(\mathbf{T}\mathbf{T}^T)^{-1}\}^{-1}. \tag{3}$$

By maximizing this FOM, we can effectively meet requirements (i) and (ii).

In a typical photonic structure containing random scatterers, each mode couples to many output channels, resulting in an unstructured transmission matrix (Figure 1, middle panel). For this case, the FOM is not close to its maximum, and the spectral reconstruction is relatively sensitive to measurement noise. However, after optimizing the structure using the above FOM, the transmission matrix will acquire a banded structure, which is predicted to give more robust and accurate spectral reconstruction (Figure 1, bottom panel). Note that the optimized transmission matrix is *not* approximately diagonal; the populated elements cluster around a non-diagonal line stretching between two corners of the non-square matrix. Physically, this means that the input signal at each frequency is directed to one or a small number of adjacent output ports, distributed evenly over the operating frequency range.

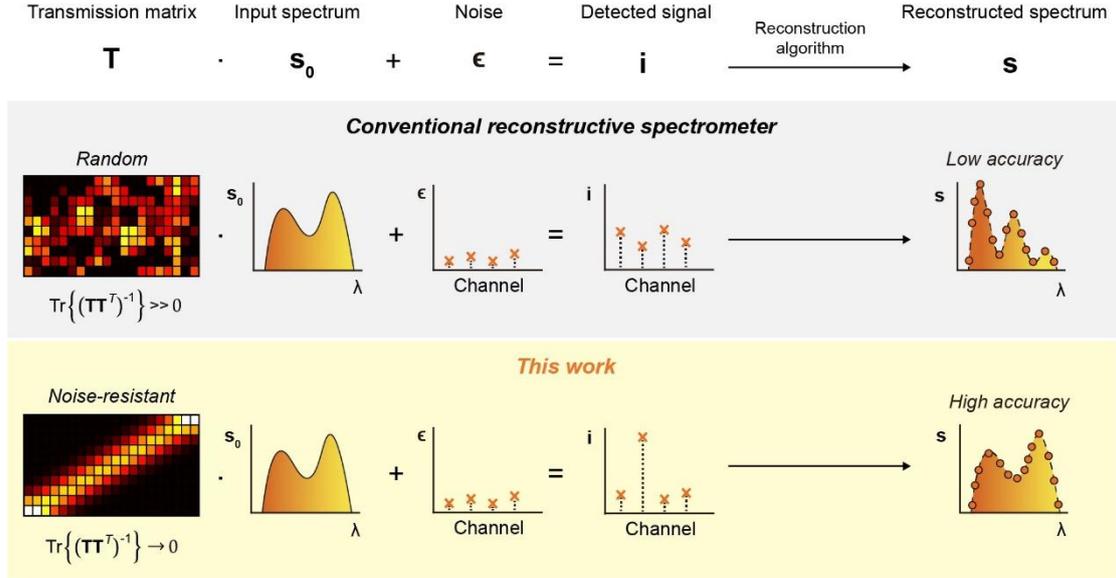

**Figure 1. Principle of noise-resistant reconstructive spectrometers.** Top panel: A reconstructive spectrometer has a spectral transmission matrix **T**. Applying **T** to an unknown input spectrum $s_0$, and adding measurement noise $\epsilon$, yields the measured intensity **i**. Using the reconstruction algorithm, we obtain the reconstructed spectrum **s**. Middle panel: for a conventional reconstructive spectrometer, **T** has a random structure, as each mode couples indiscriminately to various output ports. This increases the noise-sensitivity of the spectral reconstruction process. Bottom panel: an optimized structure has a **T** matrix with a banded structure, such that each input frequency in the operating range couples to a few output ports. This increases the robustness and accuracy of the spectral reconstruction.

*On-chip spectrometer design*

As a proof of concept for the above theory, we use inverse design to develop an on-chip spectrometer featuring the optimized banded transmission matrix described above. The spectrometer utilizes a silicon-on-insulator (SOI) platform and features a $13\ \mu m \times 13\ \mu m$ inverse designed scattering region, along with one input and 15 output waveguides. Each waveguide is 0.5 μm in width and 1.5 μm in height, and supports a single TM mode across the wavelength range of 3.60-3.82 μm. This range is divided into 85 equally spaced discrete wavelength channels, by which the spectral transmission matrix is defined. An adjoint-method inverse design procedure[30,31] is used to determine the permittivity distribution in the scattering region so that the transmission matrix closely matches the target one. The resulting structure is shown in Figure 2**a**; further design details are provided in the Supplementary Information.

The functionality of the designed spectrometer is validated through FDTD simulations, with results shown in Figure 2**b**: The upper panel displays the field distributions under monochromatic input at a few representative wavelengths. For comparison, some corresponding cavity eigenmodes are presented in the lower panel. The cavity supports around 40 eigenmodes within the target wavelength range, with Q-factors ranging from several tens to several hundred (see Supplementary information). Through inverse design, certain of these high-Q eigenmodes outcouple selectively to one or a few output ports (as for the other eigenmodes, they couple to many output ports or decay by side scattering). As a result, each output port exhibits a Lorentzian-like transmission profile (Figure 2**c**). When exciting the system at a wavelength between two adjacent transmission peaks, we are able to observe the output shifting from one port to another, as shown in the middle plot in the upper panel of Figure 2**b**. The transmission spectrum for each channel has a quality factor of 200. These high-Q quasinormal modes with extended lifetimes increase the effective optical path length to more than three times that of the random spectrometers (see Supplementary Information).

To evaluate the distinctiveness of this spectrometer's transmission matrix, we compare its FOM with those of 5,000 random structures. These random structures have the same footprint and input/output configuration as the designed spectrometer, except that the scattering region contains randomly distributed air holes (see Supplementary Information). The distribution of their FOMs is shown in Figure 2**d**. The FOM for our design is 0.12, somewhat lower than the theoretical maximum of 0.38 but also two orders of magnitude above the highest FOM among the 5,000 structures in the random sample.

We then use our design, along with the random spectrometer with the highest FOM, to reconstruct 10,000 Gaussian spectra with randomly selected bandwidths ranging from 40 to 200 nm and center wavelengths between 3.60 and 3.82 μm. The spectra are numerically input to the respective transmission matrices, and Gaussian noise is added to the output channels, with noise levels quantified by the standard deviation $\sigma$. The mean reconstruction MSE across 10,000 spectra at various noise levels from 0 to 20% is shown in Figure 2**e**. For both spectrometers, the mean MSE increases with increasing noise levels. However, the mean MSE of our inverse-designed spectrometer is more than one order of magnitude lower than that of the optimal random spectrometer across all noise levels, providing a clear indicator of superior robustness against noise.

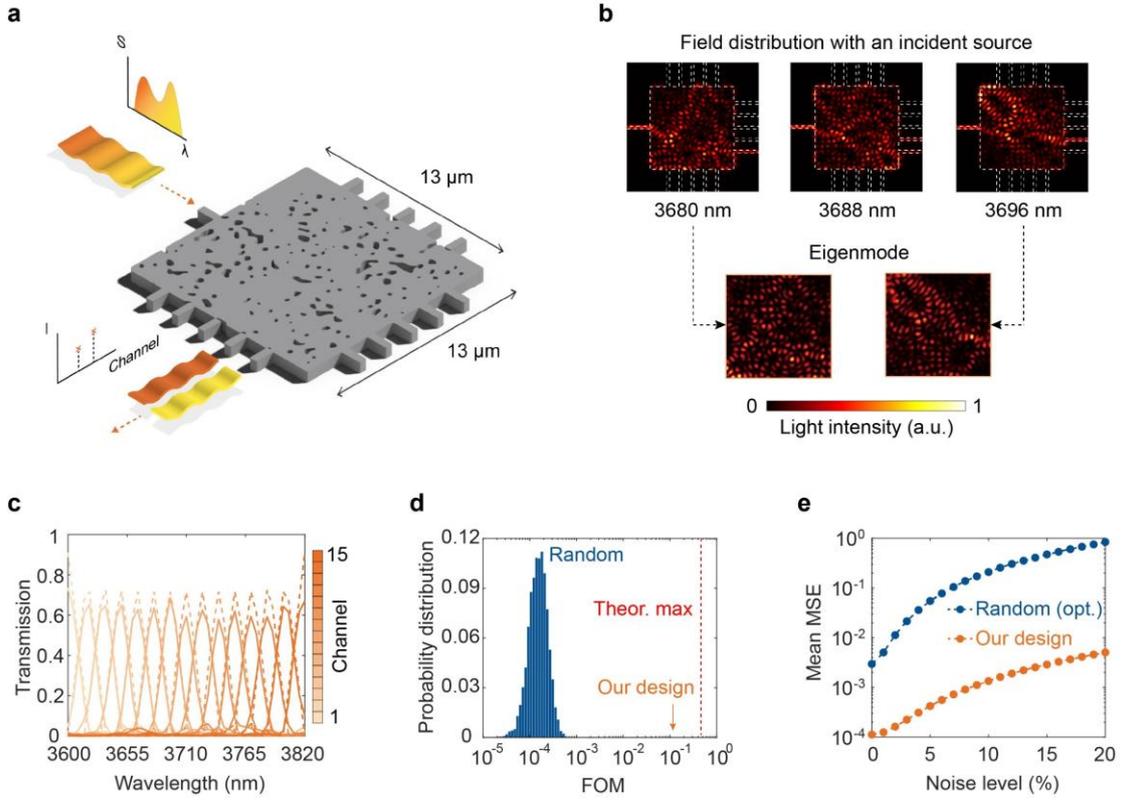

**Figure 2. Schematic of the inverse-designed spectrometer.** (a) Geometry of the inverse-designed spectrometer. (b) Top: Field distributions under monochromatic input at distinct wavelengths. Bottom: Field distributions of the corresponding cavity eigenmodes. (c) Simulated transmission at different output channels of the inverse-designed spectrometer (solid lines). The target transmission matrix is plotted for reference (dashed lines). (d) Comparison of the FOM, defined in Equation (3), between the inverse-designed spectrometer and random spectrometers. The red dashed line denotes the theoretical upper bound of FOM, $n/m^2$. € Mean reconstruction MSE across 10,000 spectra at various noise levels. The orange and blue dots represent the results of inverse-designed spectrometer and optimal random spectrometer, respectively.

*Sample fabrication and characterization*

The inverse-designed spectrometer is fabricated using standard nanofabrication processes (see Methods). A scanning electron microscopy (SEM) image of the fabricated device is shown in Figure 3**a**. Its spectral transmission matrix is characterized experimentally using a mid-infrared quantum cascade laser (QCL) with tunable wavelength. During the characterization process, a blue shift in the resonance peaks is observed due to slight over-etching of the structure, resulting in the transmission peaks for the first three channels not being visible due to falling outside the operational wavelength range of the QCL source. We utilize the remaining 12 channels for spectral reconstruction within a reduced spectral window of 3.59 μm to 3.76 μm. The measured truncated transmission matrix, shown in Figure 3**b**, clearly exhibits the sought-for banded form. To characterize it, we calculate the auto-correlation function

$$C(\Delta\lambda) = \langle \frac{\langle T(\lambda,m)T(\lambda+\Delta\lambda,m)\rangle_\lambda}{\langle T(\lambda,m)\rangle_\lambda \langle T(\lambda+\Delta\lambda,m)\rangle_\lambda} - 1 \rangle_m \quad, \tag{4}$$

where $T(\lambda, m)$ is the transmission at wavelength $\lambda$ in channel $m$, and $\langle ... \rangle$ denotes averaging over wavelengths or channels. The correlation length $\delta\lambda$, defined as the required wavelength shift to reduce the correlation by half, is around 20 nm for our device (Figure 3**c**). Note that while a smaller $\delta\lambda$ indicates a higher capability to distinguish between adjacent wavelengths, it is not necessarily the same as the spectrometer's resolution since the resolution is also influenced by other factors such as the reconstruction algorithm and the number of output channels [22].

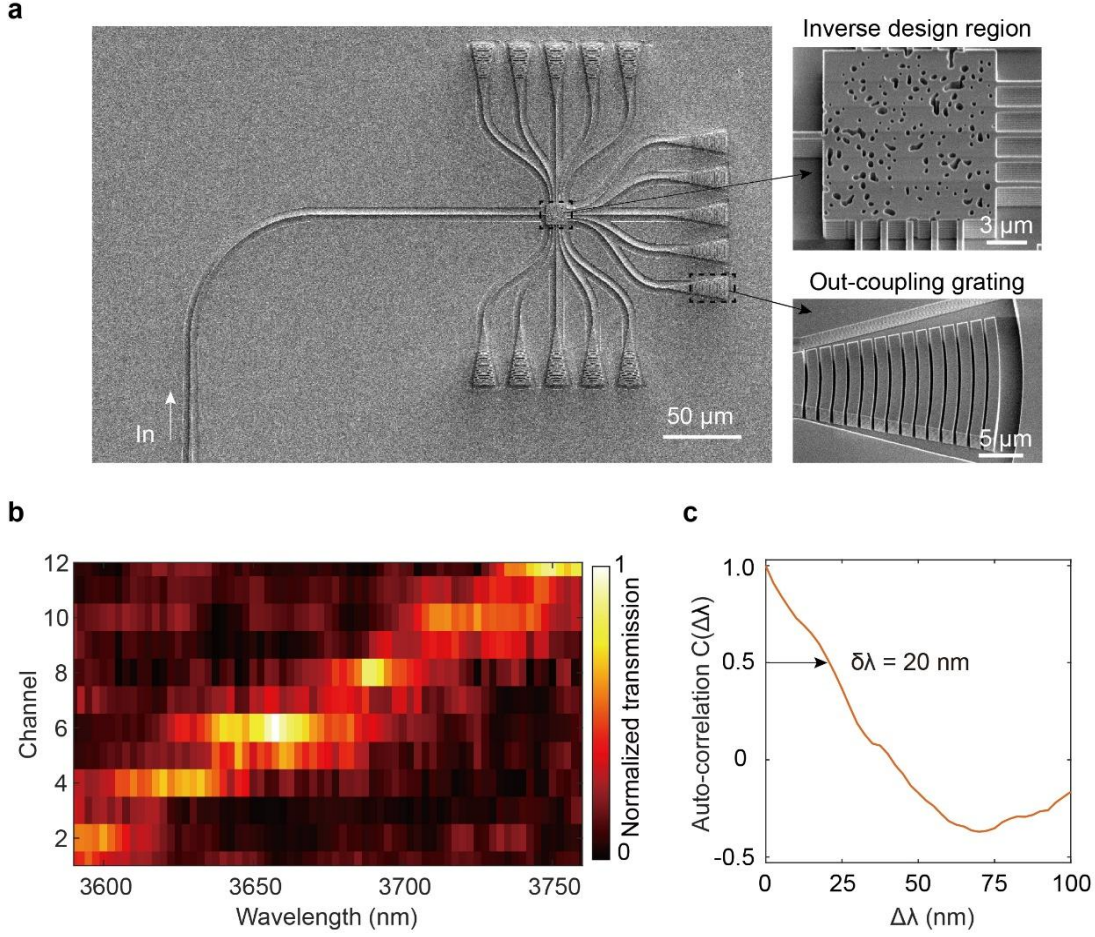

**Figure 3. Experimental characterization of the spectrometer.** (a) Scanning electron microscopy image of the fabricated spectrometer. The insets give the zoom-in view of the inverse design region and the outcoupling gratings (b) Measured transmission matrix of the spectrometer. (c) Auto-correlation function of the spectrometer.

*Spectral reconstruction*

We employ the fabricated spectrometer to reconstruct various types of spectra. In practical spectrum reconstruction tasks, prior knowledge about the spectrum can often be leveraged. For example, additional regularization terms beyond the pseudoinverse can be used when solving underdetermined problems. Here, we reconstruct the spectrum $\mathbf{s}'$ by solving the composite minimization problem

$$\mathbf{s}' = \mathrm{argmin}_S\{\|\mathbf{i} - \mathbf{Ts}\|_2^2 + \lambda_1 \|\mathbf{s}\|_1 + \lambda_2 \|\mathbf{s}\|_2^2 + \lambda_3 \|\mathbf{Ds}\|_2^2\} \tag{5}$$

where $\mathbf{D}$ is a difference operator that computes the derivative of $\mathbf{s}$. By adjusting the regularization coefficients $\lambda_1$ - $\lambda_3$, we incorporate prior knowledge about the incident spectra. Note that the choice of regularization is made post-measurement, requiring no additional experimental effort.

To quantify the noise level experimentally, we record the output intensities multiple times under identical input conditions. The relative deviation is calculated as $\Delta I_i = (I_i - \bar{I})/\bar{I}$, where $I_i$ is the output intensity at one port during the *i*-th measurement, and $\bar{I}$ is its average value. As shown in Figure 4**a**, the distribution of these relative deviations shows a Gaussian profile. This aligns with the expectation that common noise sources in experiment, such as shot noise and detector noise are approximately Gaussian. The noise level, quantified by the coefficient of variation of the intensity, is 7.9%. Based on our previous simulation studies, a more than one-order-of-magnitude reduction in reconstruction MSE was predicted compared to random spectrometers. Note that we adopted a low input power for the noisy spectral reconstruction to demonstrate the noise-resistant performance of our spectrometer. In experiment, the noise level of the output signal depends on its intensity, with higher intensities typically indicating lower noise levels (see Supplementary Information for details).

Using these noisy single-shot intensity measurements, we assess the spectrometer's ability to resolve single narrowband peaks. The reconstruction results are displayed in Figure 4**b**. The spectrometer accurately recovers the individual peak positions across the operational wavelength. Next, we evaluate the spectral resolution by synthesizing the output intensity of two closely spaced wavelengths. The reconstruction result indicates a resolution of approximately 10 nm, as shown in Figure 4**c**. Finally, we test the spectrometer's ability to reconstruct broadband spectra. We utilize a broadband supercontinuum laser in combination with a filter to generate a broadband spectrum. This signal is coupled into the spectrometer, and the reconstruction results are compared with measurements taken from a commercial FTIR spectrometer (Vertex 70, Bruker). The MSE of the broadband reconstruction is 0.005, indicating our spectrometer can reconstruct spectra accurately under noise in experiment, as illustrated in Figure 4**d**.

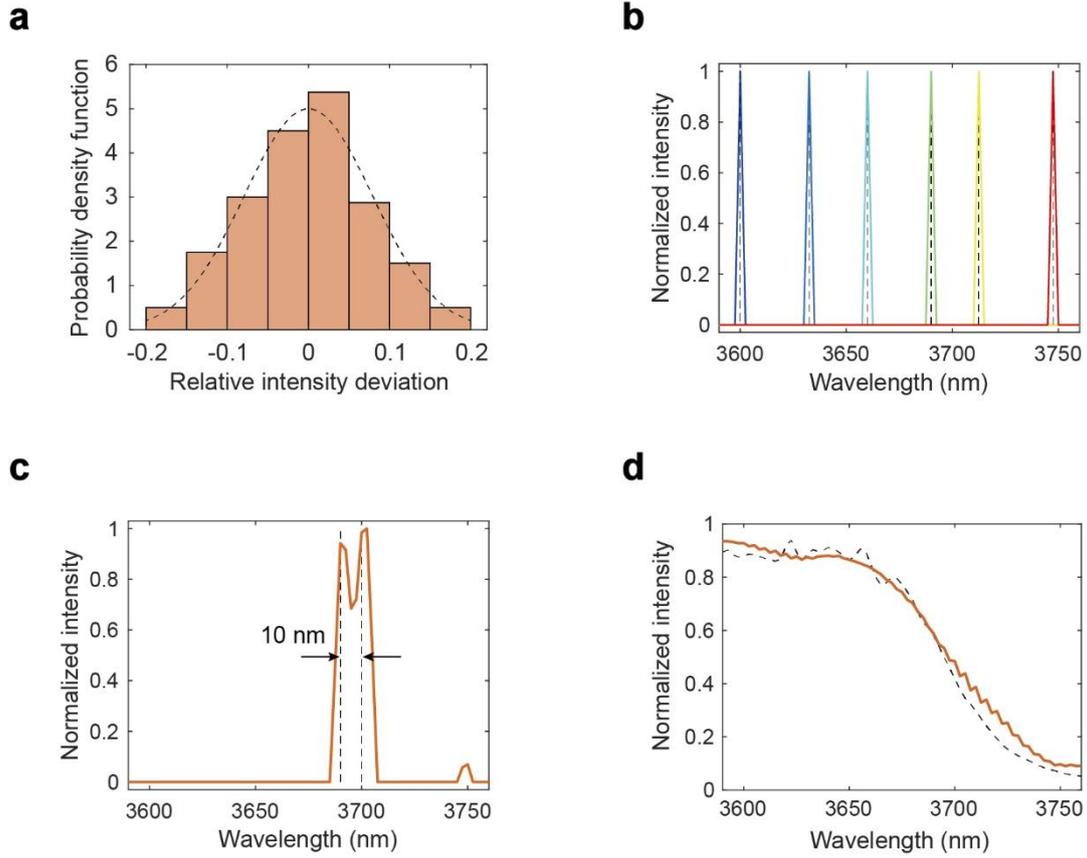

**Figure 4. Spectra reconstruction performance.** (a) Distribution of relative intensity deviation. The measurement noise is evaluated by the coefficient of variance of the intensity, which is 7.9%. The dashed line shows a Gaussian function with a standard deviation of 7.9%. (b) Reconstruction of single-peak narrowband spectra. The corresponding center wavelengths are marked by the black dashed lines. (c) Reconstruction of a dual-peak spectrum. The two narrow peaks are separated by 10 nm. (d) Reconstruction of a broadband spectrum generated by a mid-infrared supercontinuum laser combined with a filter. The dashed line is the ground truth, and the solid line shows the reconstructed spectrum.

To illustrate the advancements achieved in this work, we compare our work with existing on-chip spectrometers based on Fourier transform [18,19,32-34], narrowband filtering [13,14], dispersive components [15,17], and reconstruction [20-22,25,26,35-41], as shown in Figure 5. The key parameters of an on-chip spectrometer include its resolution $R$, bandwidth $BW$, area footprint $A$, and central working wavelength $\lambda_c$.

A normalized resolution, $R_n = R/\lambda_c$, is used to compare the resolving capabilities of spectrometers operating across different spectral bands. Generally, the resolution $R$ scales as $\sim \lambda_c^2/L_{\text{eff}}$, where $L_{\text{eff}}$ is the spectrometer's effective optical path length. In diffusive media, the path length enhancement ($L_{\text{eff}} \sim L^2/\ell_t$, as introduced earlier) causes the resolution to scale as $\sim \ell_t/A_n$, where $A_n = A/\lambda_c^2$ is the normalized footprint. Consequently, the product $R_n \cdot A_n$ reduces to the normalized transport mean free path $\sim \ell_t/\lambda_c$, with smaller values indicating an enhanced ability to scatter light and differentiate between adjacent wavelengths based on scattering patterns. The normalized resolution-footprint product, $RFP = R_n \cdot A_n$ (lower values are better), helps evaluate this resolution relative to the device's area footprint.

We can use the normalized bandwidth, $BW_n = BW/\lambda_c$, to evaluate the detectable spectral range of different spectrometers. On-chip integration necessitates spectrometers to operate across a broad band while having a compact size. Therefore, the normalized bandwidth-to-footprint ratio, $BFR = BW_n/A_n$ (higher values are better), quantifies the spectral detection capacity per unit footprint.

Among reported on-chip spectrometers, those based on Fourier transform typically require a large optical path difference and small sampling interval to ensure high resolution and broad bandwidth; narrowband filtering spectrometers require the integration of numerous high-Q resonators; dispersive spectrometers necessitate dispersive components along with extended optical paths. Consequently, these designs often involve a trade-off among resolution, bandwidth, and footprint. Reconstructive spectrometers, such as those based on disordered structures, can achieve improved performance by leveraging the randomness of the transmission matrix to ensure low correlation between adjacent channels over a wide spectral range. Nevertheless, they still require a moderate footprint, usually exceeding tens of wavelengths. Compared to current works, our wavelength-scale on-chip spectrometer demonstrates a superiority in terms of miniaturization, with the normalized $BFR$ exceeding $3 \times 10^{-3}$ and $RFP$ smaller than $4 \times 10^{-2}$, see Figure 5**a**.

As an alternative comparison, we plot the normalized footprints $A_n$ and the bandwidth-to-resolution ratios (BRR) in Figure 5**b**. The normalized footprint of our design (i.e., $A_n = 12.5$) is an order of magnitude smaller than the on-chip spectrometers reported to date. Despite its compact size, the spectral channel density of our design, evaluated by BRR per unit normalized footprint, is 1.4, exceeding most of current works that lie below the dashed line in Figure 5**b**.

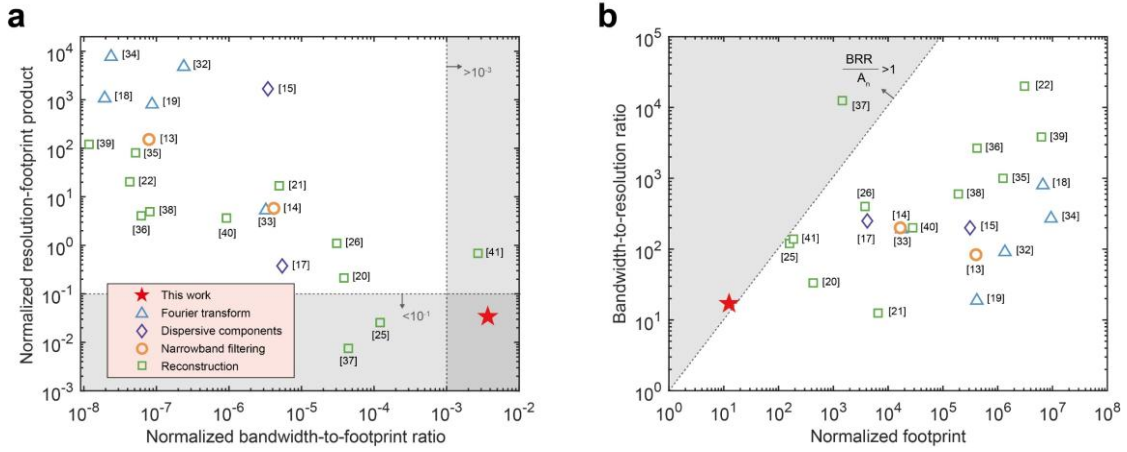

**Figure 5. Miniaturized spectrometers performance comparison.** (a) Comparison between the present device and other reported on-chip spectrometers [13-15, 17-22, 25, 26, 32-41] in terms of the resolution-footprint product and the bandwidth-to-footprint ratio. Both axes have been normalized to the central working wavelength of the spectrometers, $\lambda_c$. (b) Comparison in terms of normalized footprint and the bandwidth-to-resolution ratio (BRR).

**Conclusion**

We have proposed and experimentally demonstrated a wavelength-scale on-chip spectrometer based on silicon nanophotonic platform, achieving exceptional robustness to experimental noise alongside ultra-compact integration. Leveraging a theoretically-guided inverse design methodology, our device achieves a spectral resolution of 10 nm across the 3.59-3.76 μm wavelength range, enabling the detection of numerous molecular fingerprints in the mid-infrared region. The compact lateral footprint is only 3.5 times the operating wavelength, an order of magnitude smaller than previous reported on-chip spectrometers. Moreover, the device performs well under high noise conditions (7.9% intensity noise in our experiments), indicating its suitability for integration into chips with low-power light sources. We note that a recent theoretical work on noise-robust spectral reconstruction was posted while this manuscript was under preparation[42]. The spectral resolution can be further improved by increasing the number of detection channels. For example, utilizing the thermal-optical effect of silicon[43], the transmission matrix can be dynamically tuned by adjusting the temperature of the device, allowing for greater device flexibility and precision in spectral reconstruction. Additionally, considering the wide transparent window of silicon, the operation wavelength range of the spectrometer could be extended from 1.1 μm to 8 μm, covering wide near-infrared and mid-infrared regions, by removing the buried oxide layer to minimize the absorption loss [44]. The proposed design strategy is broadly applicable and can be extended to a variety of material platforms, including aluminum nitride, silicon nitride, lithium niobate, chalcogenide glasses, germanium, high-resistivity silicon, and polymers, enabling spectral coverage from the ultraviolet to the Terahertz. Besides, full on-chip spectral detection can be achieved by monolithically integrating photodetectors with the spectrometer, paving the way for fully self-contained spectroscopic systems. These advances highlight a versatile and scalable approach to high-resolution, noise-robust spectral sensing at a wavelength-scale size. The demonstrated platform holds strong promise for a wide range of applications in chemical analysis, biosensing, material science, environmental monitoring, and lab-on-chip photonics.

**Methods**

Sample fabrication: A 400-nm-thick resist (ZEP 520A) was spin coated on the silicon-on-insulator substrate and baked for 2 min at 180 °C. The spectrometer pattern was obtained by electron-beam lithography (EBL). The development was performed in the developer (ZED N50) for 2min followed by rinsing in isopropanol for 40s. After that, the pattern was etched using deep reactive ion etching with the remaining photoresist removed by oxygen plasma ashing.

Optical characterization: A Mid-IR tunable laser was used for characterizing the transmission matrix of the spectrometer. The output intensity was recorded by an IR camera (FLIR A6703) with the detection wavelength range of 3-5 μm.

Simulations: The transmission spectrum and the electric field distribution were calculated using the software Lumerical FDTD.

**Author contributions**

**Competing interests**

The authors declare no competing interests.

**Acknowledgement:**

Q.J.W. acknowledges the National Research Foundation Singapore program (grant nos. NRF-CRP22-2019-0007, NRF-CRP29-2022-0003, and NRF-MSG-2023-0002) and the A*STAR grant (grant nos. M22K2c0080, R23I0IR041, and M23M2b0056). Y.C. acknowledges the National Research Foundation Singapore program (grant nos. NRF-NRFI08-2022-0001).

**Supplementary Text**

**S1. Design principle of a noise-resistant transmission matrix**

For reconstructive spectrometers, the reconstruction process is to solve a linear problem, as defined by Equation 1 in the main text:

$$\mathbf{i} = \mathbf{i_0} + \boldsymbol{\epsilon} = \mathbf{T}\mathbf{s_0} + \boldsymbol{\epsilon}. \tag{S1}$$

The discretized spectrum $\mathbf{s_0}$ has dimension $n$, while the number of output channels (and thus the dimensions of $\mathbf{i}$, $\mathbf{i_0}$, and $\boldsymbol{\epsilon}$) is $m$. The transmission matrix $\mathbf{T}$ is of shape $m \times n$ and is assumed to be full rank. To minimize the footprint of the spectrometer, it is typically required that $m \ll n$, making the problem underdetermined. We employ the pseudoinverse to reconstruct the spectrum. Assuming the spectrum can be exactly reconstructed via the pseudoinverse, i.e., $\mathbf{s_0} = \mathbf{T}^+\mathbf{i_0}$, then:

$$\mathbf{s} = \mathbf{T}^+\mathbf{i} = \mathbf{s_0} + \mathbf{T}^+\boldsymbol{\epsilon}. \tag{S2}$$

The squared error (SE) of the reconstruction is given by:

$$\text{SE} = \|\mathbf{s} - \mathbf{s_0}\|^2, \tag{S3}$$

$$= \boldsymbol{\epsilon}^T(\mathbf{T}^+)^T\mathbf{T}^+\boldsymbol{\epsilon}, \tag{S4}$$

$$= \boldsymbol{\epsilon}^T(\mathbf{T}\mathbf{T}^T)^+\boldsymbol{\epsilon}, \tag{S5}$$

$$= \boldsymbol{\epsilon}^T(\mathbf{T}\mathbf{T}^T)^{-1}\boldsymbol{\epsilon}, \tag{S6}$$

where the following properties of the pseudoinverse are used: $(\mathbf{T}^+)^T = (\mathbf{T}^T)^+$, $(\mathbf{T}\mathbf{T}^T)^+ = (\mathbf{T}^T)^+\mathbf{T}^+$, and $(\mathbf{T}\mathbf{T}^T)^+ = (\mathbf{T}\mathbf{T}^T)^{-1}$. Assuming elements of the noise vector $\boldsymbol{\epsilon}$ are independently Gaussian-distributed with zero mean and standard deviation $\sigma$, the mean squared error (MSE) is a standard result for matrix quadratic forms, which is given by:

$$\text{MSE} = \mathbb{E}\{\|\mathbf{s} - \mathbf{s_0}\|_2^2\} = \sigma^2 \text{Tr}\{(\mathbf{T}\mathbf{T}^T)^{-1}\}, \tag{S7}$$

where $\mathbb{E}\{\ldots\}$ denotes the expectation value. To examine how the MSE is bounded, we define $\mathbf{T}^T = (T_1, T_2, \ldots, T_m)$, where $T_i$ represents the transmission function of the $i$-th channel. The cross-correlation matrix is given by:

$$\mathbf{T}\mathbf{T}^T = \begin{bmatrix} \|T_1\|^2 & \cdots & T_1^T \cdot T_m \\ \vdots & \ddots & \vdots \\ T_m^T \cdot T_1 & \cdots & \|T_m\|^2 \end{bmatrix}. \tag{S8}$$

For an ideal spectrometer, the incident light has unity transmission across the detected spectral range, i.e., $\sum T_i^T = (1,1,\ldots,1)$, and the element sum of the cross-correlation matrix is:

$$\sum T_i^T \cdot T_j = \left(\sum T_i^T\right) \cdot \left(\sum T_j\right) = n. \tag{S9}$$

This imposes an upper bound on the trace of the cross-correlation matrix, which equals the sum of

its eigenvalues and can be computed as the sum of its diagonal elements:

$$\text{Tr}\{\mathbf{T}\mathbf{T}^T\} = \sum \lambda_i = \sum \|T_i\|^2 \leq \sum T_i^T \cdot T_j = n, \tag{S10}$$

where $\lambda_i$ denotes the eigenvalues of the cross-correlation matrix. Equality holds when $\sum_{i \neq j} T_i^T \cdot T_j = 0$, i.e., the transmission functions are uncorrelated across different channels. Combining Equation (S10) with the HM–AM inequality gives:

$$\sum \frac{1}{\lambda_i} \geq \frac{m^2}{\sum \lambda_i} \geq \frac{m^2}{n}. \tag{S11}$$

The left-hand side of this inequality equals $\text{Tr}\{(\mathbf{T}\mathbf{T}^T)^{-1}\}$, which determines the MSE in Equation (S7). Equality in the HM–AM inequality holds when all eigenvalues are equal, i.e., $\lambda_i = n/m$, as inferred from Equation (S10). The cross-correlation matrix satisfying these conditions is:

$$\mathbf{T}\mathbf{T}^T = \frac{n}{m}\mathbb{I}, \tag{S12}$$

where $\mathbb{I}$ denotes the identity matrix. This leads to the MSE lower bound presented in the main text:

$$\text{MSE} \geq \frac{\sigma^2 m^2}{n}. \tag{S13}$$

There exist infinitely many transmission matrices that satisfy Equation (S12) and thus achieve the MSE lower bound. However, the transmission matrix should also meet additional physical constraints in a spectrometer. A key requirement is the smoothness of the transmission functions. We numerically optimize the transmission matrix by maximizing the FOM defined in the main text while enforcing smoothness on the transmission functions. The optimization yields a series of Lorentzian-like transmission functions, confirming that quasinormal modes are robust to noise in reconstructive spectrometers.

## S2. Details on inverse design

The inverse design process is based on a standard adjoint-method optimization procedure[1] with the steps stated as follows.

Step 1: The permittivity within the inverse-designed region is set to be that of silicon, i.e., $\varepsilon(x,y) = \varepsilon_{Si}$. The transmission matrix of this initial structure is simulated by FDTD Solutions, and the performance of the spectrometer is evaluated by a cost function, which is defined as the sum of absolute differences between the target transmission matrix (the numerically optimized transmission matrix mentioned in the main text) and the simulated transmission matrix.

Step 2: The gradient of the cost function with respect to the permittivity distribution is evaluated by performing an adjoint FDTD simulation, based on which a permittivity perturbation $\Delta\varepsilon(x,y)$ is introduced within the optimization region.

Step 3: Step 2 is applied iteratively until convergence is achieved. Finally, the permittivity distribution is binarized to obtain the optimized structure.

To proceed the optimization process, we first perform 2D optimization. The optimization undergoes 616 interactions before convergence. For the final 2D optimized structure, the total transmission of all channels is 90% on average, and the losses mainly arise from in-plane scattering (Figure S1 (a)). After the 2D optimization, a 3D optimization is performed, which takes the out-of-plane scattering into consideration, and are thereby closer to the real case. The permittivity distribution obtained from 2D optimization is chosen as the initial guess for 3D optimization. The 3D optimization undergoes 50 interactions before convergence. The total transmission of the 3D optimized structure is around 60%, and the losses arise from both in-plane and out-of-plane scattering (Figure S1 (b)).

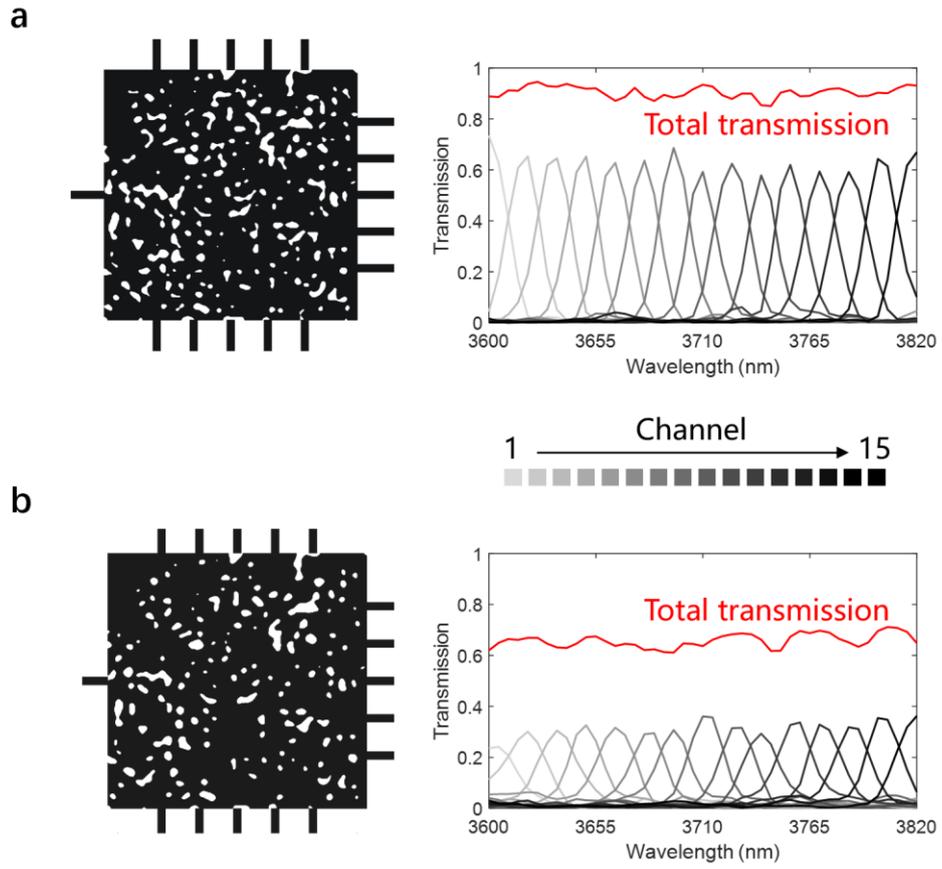

**Figure S1. Inverse design of the spectrometer.** (a) Geometry and transmission matrix of the 2D inversely designed structure. (b) Geometry and transmission matrix of the 3D inversely designed structure.

## S3. Eigenfrequencies calculation

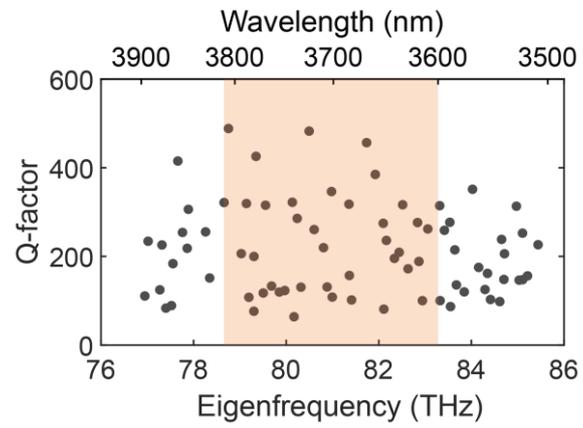

**Figure S2. Eigenfrequencies of the inverse-designed spectrometer.** The Q-factors of eigenmodes range from several tens to several hundred. The colored region denotes the target wavelength range from 3600 nm to 3820 nm.

## S4. Spectrometers based on random structures

For spectrometers based on random structures, they have the same configuration as the inverse-designed spectrometer, except that the scattering region contains randomly distributed air holes. The geometry of one representative random spectrometer is illustrated in Figure S3 (a). Each air hole has a radius of $r = 100$ nm, and its position is determined by the random function in Matlab. Five types of random structures are generated, differentiated by the minimum gap between adjacent air holes: 100 nm, 150 nm, 200 nm, 250 nm, and 300 nm, respectively. For each type, 1,000 structures are created, resulting in a total of 5,000 random structures. Their transmission matrices are computed using FDTD Solutions. The transmission matrix of the optimum random structure is shown in Figure S3 (b), which achieves the maximum FOM of $5.7 \times 10^{-4}$ among all random structures.

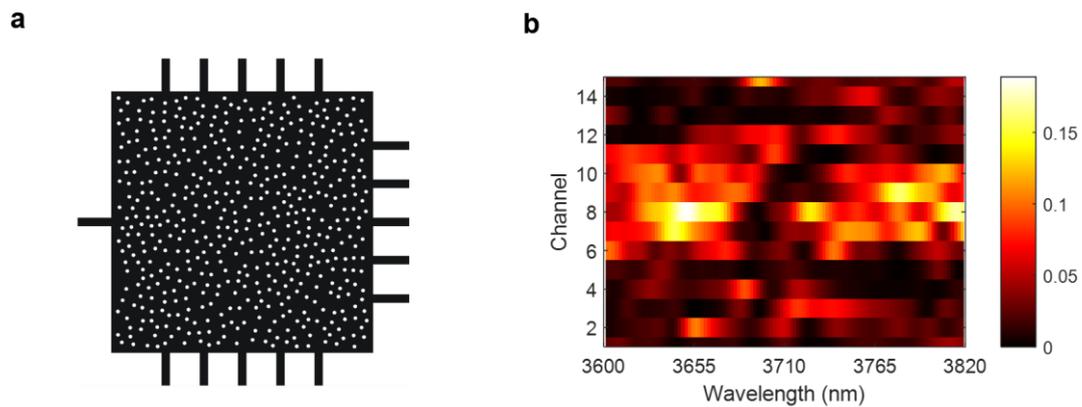

**Figure S3. Spectrometer based on random structures.** (a) Geometry of one representative random spectrometer. (b) The transmission matrix of the optimum random spectrometer.

## S5. Evaluation of optical path length

For random spectrometers, the transport mean free path $\ell_t$ is computed using the relation $T = \frac{\ell_t}{L}$, where $T$ is the total transmittance averaged across the target wavelength range, and $L$ is the lateral size of the spectrometer ($L = 13$ μm). In the simulation, we calculate the transmission of 1,000 random structures, and get an average value of $T = 0.67$. Then, we get the transport mean free path in random spectrometers as $\ell_t = 8.71$ μm. The effective optical path length is estimated to be $l_{\text{eff}} \sim \frac{L}{T} = 19$ μm.

For our inverse-designed spectrometer, each output has a Lorentzian function, whose quality factor is expressed as

$$Q = \frac{\omega_r}{2} \cdot \tau = \frac{\pi c}{\lambda_r} \cdot \tau$$

where $\omega_c$ ($\lambda_r$) is the resonant central frequency (wavelength), $c$ is the vacuum light speed, and $\tau$ is the mode lifetime. The effective optical path length $l_{\text{eff}}$ is expressed as

$$l_{\text{eff}} = \frac{c}{n_{\text{eff}}} \cdot \tau$$

where $n_{\text{eff}}$ is the effective refractive index. The quality factor, obtained from the simulation results, is $Q = 200$. Therefore, the effective optical path length for the inverse-designed spectrometer is calculated to be $l_{\text{eff}} = 70$ μm, which is more than three times longer than that of the random spectrometers.

## S6. Experimental setup

The experimental setup for transmission matrix characterization and spectra reconstruction is depicted in Figure S4 (a). For transmission matrix characterization, a mid-infrared quantum cascade laser (QCL) serves as the input light source. The laser is coupled into the incident waveguide through an edge taper, and its output wavelength is scanned from 3590 nm to 3760 nm with a step size of 2.5 nm. The output light from each channel is directed by a waveguide and scattered by second-order Bragg gratings. The out-of-plane scattering intensity is recorded by an IR camera. The IR images of the scattering pattern at some selected wavelengths are shown in Figure S4 (b). For spectra reconstruction, the light source is replaced with a broadband supercontinuum laser combined with a filter. The output intensity at each channel is obtained from the IR image, based on which the incident spectrum can be reconstructed.

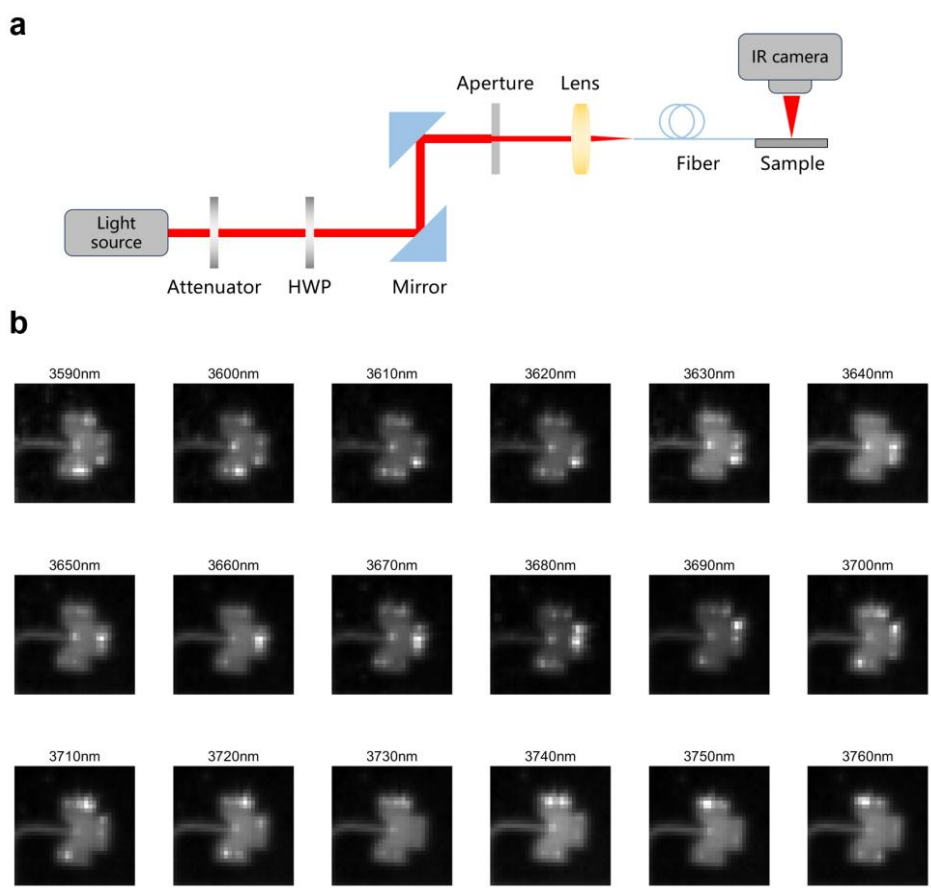

**Figure S4. Experimental results.** (a) Experimental setup for transmission matrix characterization and spectra reconstruction. HWP: half-wave plate. (b) IR images of the scattering pattern at some selected wavelengths.

## S7. Noise level evaluation in experiment

During the measurement, the incident laser is coupled into the waveguide through an edge taper, and the out-of-plane scattering intensity is recorded by an IR camera. The detected scattering power ($I_S$) is the sum of the laser contribution ($I_L$) and the background thermal emission ($I_B$, measured under zero incident illumination). Therefore, the output intensity of the spectrometer is computed as $I_L = I_S - I_B$.

To evaluate the noise level of $I_L$, we first fix the input power and wavelength of the mid-IR QCL, and record the output intensity multiple times. Then, the mean output intensity $\bar{I}_L$ is computed, and the noise level is evaluated using the coefficient of variation. Next, we increase the input power to raise $\bar{I}_L$ and repeat the measurement process. The noise levels at various $\bar{I}_L$ values are summarized in Table S1. As shown, the noise level decreases as the output intensity increases.

**Table S1. Noise level at various $\bar{I}_L$**

| $\bar{I}_L$ (a.u.) | 4.2 | 8.1 | 13.2 | 17.6 | 26.3 |
|---|---|---|---|---|---|
| Noise level | 13.4% | 7.9% | 4.4% | 3.6% | 1.6% |